\documentclass[mathleft
]{an}
\usepackage{graphicx}
\usepackage{times}
\begin{document}

% The following seven commands are intended for editorial usage and should be ignored by
% the author(s).
\Pagespan{789}{}% Document's page range.
% If second parameter is left empty, the last page is computed automatically.
\Yearpublication{2011}%
\Yearsubmission{2010}%
\Month{11}%
\Volume{999}%
\Issue{88}%
% \DOI{This.is/not.aDOI}%

\title{Young Exoplanet Transit Initiative (YETI)}

\title{Young Exoplanet Transit Initiative (YETI)}

\author{R. Neuh\"auser\inst{1} \thanks{Corresponding author: \email{rne@astro.uni-jena.de}}
%Example
%for footnote, note the usage of the \texttt{fnmsep}
%command as separator between institute number and footnote mark}
\and R. Errmann\inst{1}
\and A. Berndt\inst{1}
\and G. Maciejewski\inst{1,2}
\and H. Takahashi\inst{3}
\and W.P. Chen\inst{4}
\and D.P. Dimitrov\inst{5}
\and T. Pribulla\inst{6,1}
\and E.H. Nikogossian\inst{7}
\and E.L.N. Jensen\inst{8}
\and L. Marschall\inst{9}
\and Z.-Y. Wu\inst{10}
\and A. Kellerer\inst{11,12}
\and F.M. Walter\inst{13}
\and C. Brice\~no\inst{14}
\and R. Chini\inst{15,16}
\and M. Fernandez\inst{17}
\and St. Raetz\inst{1}
\and G. Torres\inst{18}
\and D.W. Latham\inst{18}
\and S.N. Quinn\inst{18}
\and A. Niedzielski\inst{2}
\and {\L}. Bukowiecki\inst{2}
\and G. Nowak\inst{2}
\and T. Tomov\inst{2}
\and K. Tachihara\inst{19,20}
\and S.C.-L. Hu\inst{4}
\and L.W. Hung\inst{4}
\and D.P. Kjurkchieva\inst{21} 
\and V.S. Radeva\inst{21}
\and B.M. Mihov\inst{5}
\and L. Slavcheva-Mihova\inst{5}
\and I.N. Bozhinova\inst{5}
\and J. Budaj\inst{6}
\and M. Va\v{n}ko\inst{6}
\and E. Kundra\inst{6}
\and \v{L}. Hamb\'alek\inst{6}
\and V. Krushevska\inst{6,22}
\and T. Movsessian\inst{7}
\and H. Harutyunyan\inst{7}
\and J.J. Downes\inst{14}
\and J. Hernandez\inst{14}
\and V.H. Hoffmeister\inst{15}
\and D.H. Cohen\inst{8}
\and I. Abel\inst{8}
\and R. Ahmad\inst{8}
\and S. Chapman\inst{8}
\and S. Eckert\inst{8}
\and J. Goodman\inst{8}
\and A. Guerard\inst{8}
\and H.M. Kim\inst{8}
\and A. Koontharana\inst{8}
\and J. Sokol\inst{8}
\and J. Trinh\inst{8}
\and Y. Wang\inst{8}
\and X. Zhou\inst{10}
\and R. Redmer\inst{23}
\and U. Kramm\inst{23}
\and N. Nettelmann\inst{23}
\and M. Mugrauer\inst{1}
\and J. Schmidt\inst{1}
\and M. Moualla\inst{1}
\and C. Ginski\inst{1}
\and C. Marka\inst{1}
\and C. Adam\inst{1}
\and M. Seeliger\inst{1}
\and S. Baar\inst{1}
\and T. Roell\inst{1}
\and T.O.B. Schmidt\inst{1}
\and L. Trepl\inst{1}
\and T. Eisenbei\ss \inst{1}
\and S. Fiedler\inst{1}
\and N. Tetzlaff\inst{1}
\and E. Schmidt\inst{1}
\and M.M. Hohle\inst{1,24}
\and M. Kitze\inst{1}
\and N. Chakrova\inst{25}
\and C. Gr\"afe\inst{1,26}
\and K. Schreyer\inst{1}
\and V.V. Hambaryan\inst{1}
%\and H. Gilbert\inst{1}
%\and F. Gie\ss ler\inst{1}
\and C.H. Broeg\inst{27}
\and J. Koppenhoefer\inst{24,28}
\and A.K. Pandey\inst{29}
}

\titlerunning{Young exoplanet transit initiative}
\authorrunning{R. Neuh\"auser}

\institute{
Astrophysikalisches Institut und Universit\"ats-Sternwarte, FSU Jena,
Schillerg\"a\ss chen 2-3, D-07745 Jena, Germany
\and 
Toru\'n Centre for Astronomy, Nicolaus Copernicus University, Gagarina 11, PL–87-100 Toru\'n, Poland
\and
Gunma Astronomical Observatory, 6860-86 Nakayama, Takayama-mura, Agatsuma-gun, Gunma 377-0702 Japan
\and
Graduate Institute of Astronomy, National Central University, Jhongli City, Taoyuan County 32001, Taiwan (R.O.C.)
\and
Institute of Astronomy and NAO, Bulg. Acad. Sc., 72 Tsarigradsko Chaussee Blvd., 1784 Sofia, Bulgaria
\and 
Astronomical Institute, Slovak Academy of Sciences, 059 60, Tatransk\'a Lomnica, Slovakia
\and
Byurakan Astrophysical Observatory, 378433 Byurakan, Armenia 
\and
Dept. of Physics and Astronomy, Swarthmore College, Swarthmore, PA 19081-1390, USA
\and
Gettysburg College Observatory, Department of Physics, 300 North Washington St., Gettysburg, PA 17325, USA
\and 
Key Laboratory of Optical Astronomy, NAO, Chinese Academy of Sciences, 20A Datun Road, Beijing 100012, China
\and
Institute of Astronomy, University of Hawaii, 640 N. A'ohoku Place, Hilo, Hawaii 96720
\and
Big Bear Solar Observatory, 40386 North Shore Lane, Big Bear City, CA 92314-9672, USA
\and
Department of Physics and Astronomy, Stony Brook University, Stony Brook, NY 11794-3800, USA
\and 
Centro de Investigaciones de Astronomia, Apartado Postal 264, Merida 5101, Venezuela
\and
Astronomisches Institut, Ruhr-Universit\"at Bochum, Universit\"atsstr. 150, D-44801 Bochum, Germany
\and
Facultad de Ciencias, Universidad Cat\'{o}lica del Norte, Antofagasta, Chile
\and
Instituto de Astrofisica de Andalucia, CSIC, Apdo. 3004, 18080 Granada, Spain
\and
Harvard-Smithsonian Center for Astrophysics, 60 Garden St., Mail Stop 20, Cambridge MA 02138, USA
\and
Joint ALMA Observatory, Alonso de C\'ordova 3107, Vitacura, Santiago, Chile
\and
National Astronomical Observatory of Japan, ALMA project office, 2-21-1 Osawa Mitaka Tokyo 181-8588 Japan
\and
Shumen University, 115 Universitetska str., 9700 Shumen, Bulgaria
\and 
Main Astronomical Observatory of NAS of Ukraine, 27 Akademika Zabolotnoho St., 03680 Kyiv, Ukraine
\and
Institut f\"ur Physik, Universit\"at Rostock, D-18051 Rostock, Germany
\and
MPI f\"ur Extraterrestrische Physik, Gie\ss enbachstra\ss e 1, D-85740 Garching, Germany
\and
Institute for Applied Physics, FSU Jena, Max-Wien Platz 1, D-07743 Jena, Germany
\and
Christian-Albrechts-Universit\"at Kiel, Leibnizstra\ss e 15, D-24098 Kiel, Germany
\and
Physikalisches Institut, University of Bern, Sidlerstra\ss e 5, CH-3012 Bern, Switzerland
\and
Universit\"ats-Sternwarte M\"unchen, Scheinerstra\ss e 1, D-81679 M\"unchen, Germany
\and
Aryabhatta Research Institute of Observational Science, Manora Peak, Naini Tal, 263 129, Uttarakhand, India
}

\received{2011 April 20}
\accepted{2011 June 15}

\keywords{star formation, planet formation, extra-solar planets}

\abstract{We present the {\em Young Exoplanet Transit Initiative} (YETI),
in which we use several 0.2 to 2.6m telescopes around the world to monitor continuously
young ($\le 100$ Myr), nearby ($\le 1$ kpc) stellar clusters mainly to detect young transiting
planets (and to study other variability phenomena on time-scales from minutes to years).
The telescope network enables us to observe the targets continuously 
for several days in order not to miss any transit.
The runs are typically one to two weeks long, about three runs per year per cluster 
in two or three subsequent years for 
about ten clusters.
There are thousands of stars detectable in each field with several 
hundred known cluster members,
e.g. in the first cluster observed, Tr-37, a typical cluster for the YETI survey,
there are at least 469 known young stars detected in YETI data 
down to R=16.5 mag with sufficient precision of 50 milli-mag rms
(5 mmag rms down to R=14.5 mag) to detect
transits, so that we can expect at least about one young transiting object in
this cluster. If we observe $\sim 10$ similar clusters, we can expect to detect
$\sim 10$ young transiting planets with radius determinations.
The precision given above is for a typical telescope of the YETI network, 
namely the 60/90-cm Jena telescope (similar brightness limit, namely within $\pm 1$ mag,
for the others)
so that planetary transits can be detected. 
For targets with a periodic transit-like light curve, we obtain
spectroscopy to ensure that the star is young and that
the transiting object can be sub-stellar; then, we obtain Adaptive Optics infrared images
and spectra, to exclude other bright eclipsing stars in the (larger) optical PSF; 
we carry out other observations as needed to rule out other false positive scenarios; 
finally, we also perform spectroscopy 
to determine the mass of the transiting companion.
For planets with mass and radius determinations,
we can calculate the mean density and probe the internal structure. 
We aim to constrain planet formation models and their time-scales
by discovering planets younger than $\sim 100$ Myr and 
determining not only their orbital parameters,
but also measuring their true masses and radii, 
which is possible so far only by the transit method.
Here, we present an overview and first results.}

\maketitle

\section{Introduction: Extrasolar planets}

Beginning with the discovery of planets around a neutron star (Wolszczan \& Frail 1992; Wolszczan 1994) 
and around normal stars (Latham et al. 1989; Mayor \& Queloz 1995; Marcy \& Butler 1996), 
it has become possible to study planetary systems and their formation outside the Solar System.

The most successful of the detection methods (the spectroscopic or radial velocity (RV) technique) 
yields only a lower limit $m \cdot \sin i$ on the mass $m$ of the companions, because the orbital
inclination $i$ is unknown. RV companions could be planets, brown dwarfs,
or even low-mass stars. Combined with other observational techniques
(e.g. astrometry, Benedict et al. 2002), one can
determine the orbital inclination $i$ and the true mass $m$. 

The very existence of a transit requires that $\sin i$ is close to 1,
hence it confirms an RV planet candidate to be really below some upper
mass limit of planets (see below for the definition of planets). 
The transit technique can then even give the planetary radius and,
together with the mass (from RV and transit data), also the mean density (e.g. Torres, Winn, Holman 2008).
Modeling can then constrain the chemical composition and the mass of a solid 
core (e.g. Guillot et al. 2006; Burrows et al. 2007; Nettelmann et al. 2010).
The first transiting extrasolar planet identified was the planet candidate HD 209458b found by 
RV by Mazeh et al. (2000) and confirmed to be a planet with 0.7 Jup masses by transit 
observations (Charbonneau et al. 2000; Henry et al. 2000);
it is also the first RV planet candidate confirmed by another technique.
For transiting planets, one can also obtain spectral information by
transmission spectroscopy (e.g. Charbonneau et al. 2002).
One can also indirectly determine the brightness of the
planet by detecting the secondary eclipse (e.g. Deming et al. 2005).
The Rossiter Mc- Laughlin effect can provide information about spin axis-orbital plane
alignment and, hence, dynamics in the 
system (e.g. Queloz et al. 2000; Triaud et al. 2010; Winn et al. 2010).

The direct imaging technique can detect planets or candidates at wide separations from the star,
for which one can then often also take spectra, e.g. GQ Lup b (Neuh\"auser et al. 
2005) or CT Cha b (Schmidt et al. 2008).
The mass is difficult to determine and model-dependent,
so that all or most directly imaged planets (or planet candidates) can be either
planets or brown dwarfs. The upper mass limit of planets is also not yet defined
and can be either the Deuterium burning mass limit ($\sim 13$~M$_{\rm Jup}$, Burrows et al. 1997)
or the mass range of the brown dwarf desert ($\sim 35$~M$_{\rm Jup}$, Grether \& Lineweaver 2006).

Some additional planets or planet candidates detected by microlensing or timing have not yet
been confirmed (see, e.g, exoplanet.eu for references).

The detectability of a planet depends on its mass, radius, or luminosity
(depending on the technique) {\em relative} to the host star.
Since all detection techniques are biased towards more massive (or larger) planets,
it is not surprising that Earth-mass planets have not yet been discovered.
Low-mass planets in the so-called habitable zone can possibly be detected by
the RV technique around low-mass stars like M dwarfs (e.g. GJ 1214 by Charbonneau et al. 2009).
The RV and transit techniques are also biased towards close-in planets,
while direct imaging is biased towards wide separations.
The RV technique has led to the
discovery of $\sim 500$ planet candidates, of which some $20~\%$ are confirmed by 
either the astrometry or transit technique (see e.g. exoplanet.eu for updates). 
The latter method has detected $\sim 100$ planets, all of which have been confirmed by 
RV data (or transit timing, see Lissauer et al. 2011). Many more transit candidates
are reported by Borucki et al. (2011) with the Kepler satellite.

Almost 50 planet host stars are already known
to be surrounded by more than one planet (e.g. Fischer et al. 2002; Lovis et al. 2011), 
e.g. most recently Kepler-11 with six transiting planets, where the transit timing
variations (TTV) could be used to determine the masses of the planets (Lissauer et al. 2011)
instead of radial velocity follow-up;
several planet host stars are multiple stars themselves 
(Cochran et al. 1997; Mugrauer, Neuh\"auser, Mazeh 2007).
Planetary systems can also be discovered (indirectly) by
TTV (Maciejewski et al. 2010, 2011a; Holman et al. 2010).
One can also observe in several cases dust debris disks around planet host
stars, which are produced by colliding planetesimals (e.g. Krivov 2010), 
e.g. the probably planetary mass companions around the A0-type star HR 8799
discovered by Marois et al. (2008, 2010) and studied in detail by Reidemeister et al. (2009)
with a debris disk resolved by Su et al. (2010).

Stellar activity can be a problem for the RV, transit, TTV, and astrometric techniques,
but not for direct imaging.
Most of the planet host stars are main-sequence G-type stars,
few planets are also discovered around more or less massive stars
including giants (Frink et al. 2002; Niedzielski et al. 2007) and 
M-type dwarfs (Marcy et al. 1998; Charbonneau et al. 2009), respectively. 
The MEarth project is searching for transiting planets around M-type dwarfs
(see, e.g., Charbonneau et al. 2009).
Almost all the planets (and host stars) are, however, Gyr old,
so that it might be difficult to study planet formation from this sample.
Among the important overall statistical results is the fact that
many planets orbit their stars on much shorter orbits than in the
Solar System; hence, planets may migrate inwards after
formation further outwards (Goldreich \& Tremaine 1980; Lin et al. 1996),
e.g. beyond the ice line, if they did not form in-situ.
In addition, many more planets have been detected around metal-rich
stars than around metal-poor stars (e.g. 
Marcy et al. 2005; Butler et al. 2006), 
which may suggest that planets are more likely 
to form when there is a more abundant supply of dust.

Planets (or planet candidates) around pre-main sequence (PMS) stars or stars significantly younger
than 100 Myr (the maximal PMS time-scale for very low-mass stars) have been discovered
so far only with the direct imaging technique, so that the mass and, hence, 
the planetary status of those objects are model-dependent and still uncertain. 
The youngest known planetary system -
except maybe HR 8799 (Marois et al. 2008, 2010, four planets by direct imaging) - may be
WASP-10bc, where WASP-10b was discovered by the transit technique
and confirmed by RV (Christian et al. 
2009) and WASP-10c by transit timing (Maciejewski et al. 2011a),
still to be confirmed independently;
the age of WASP-10 was inferred from the 12 day rotation period using 
gyro-chronology to be only 200 to 350 Myr (Maciejewski et al. 2011a).
In addition, there were also RV surveys for planets among young,
PMS stars, e.g. Guenther \& Esposito (2007) have monitored 85 young
pre- and zero-age main sequence stars with ESO 3.6m HARPS without
planet discoveries; 
Joergens (2006) observed 12 very low-mass PMS stars 
and brown dwarfs in Cha I and found one companion
with $25 \pm 7$ to $31 \pm 8$ Jup mass lower mass limit around Cha H$\alpha$ 8
(Joergens \& M\"uller 2007; Joergens, M\"uller, Reffert 2010);
and Setiawan et al. (2008) monitored young stars including
PMS stars and published an RV planet candidate 
around TW Hya, which was not confirmed by Hu{\'e}lamo et al. (2008).

To study planet formation 
(planets younger than 100 Myr, partly even younger than $\sim 10$ Myr),
measuring their ages, masses, radii, and orbital elements,
studying their internal structure) and possible secular effects
(by comparing the architecture, i.e. the number and properties
of planets including their semi-major axes, of young planetary systems
with the Solar System and other old systems), we started a project
to monitor young stellar clusters (age $\le 100$ Myr) in order to find young transiting planets
(called YETI for Young Exoplanet Transit Initiative\footnote{www.astro.uni-jena.de/YETI.html})
and also to study other variability in general in the young stars.
A first brief presentation of the project was given in Maciejewski et al. (2011b).

In this paper, we first summarize previous and/or ongoing searches for young planets in clusters
(Sect. 2) and then describe the YETI target selection criteria and list the first few clusters
being observed (Sect. 3). In Sect. 4 we present the telescope
network put together for continuous monitoring and follow-up observations. 
We mention other science projects to be studied with the same data sets in Sect. 5.
Finally, we present in Sect. 6 the YETI data reduction technique and then also 
a few preliminary YETI results from the Trumpler 37 (Tr-37) cluster observed in 2009 and 2010.

\section{Previous searches for young transit planets}

We know of two other previous and/or ongoing searches for planetary transits in young clusters:
CoRoT's survey of NGC 2264 and the MONITOR project.

The French-European satellite CoRoT with its 30 cm mirror 
continuously monitored the young cluster NCG 2264 for 24 days in March 2008,
the PI being F. Favata.
NGC 2264 is $\sim 3$ Myr old at $\sim 760$ pc (Dahm 2008).
Some first preliminary results (rotation periods of member stars, 
but no transit candidates) were presented by Favata et al. (2010).
Because of the unprecedented precision of a space telescope like CoRoT,
we will not observe NGC 2264 in the YETI survey.

The MONITOR project aims at observing ten young clusters
(1-200 Myr) also mainly to detect transiting planets (Hodgkin et al. 2006,
Aigrain et al. 2007), including h and $\chi$ Per, also potential
target clusters for the YETI survey.
So far, rotation periods of hundreds of member stars have been reported
for the clusters M34 (Irwin et al. 2006), NGC 2516 (Irwin et al. 2007),
NGC 2362 (Irwin et al. 2008a), NGC 2547 (Irwin et al. 2008b),
and M50 (Irwin et al. 2009). Results for the planet transit search
have been published so far only for NGC 2362, where no planet was found
among 475 member stars observed for a total of 100 hours spread over
18 nights within 362 days (Miller et al. 2008); the non-detection
of transiting planets was not surprising given the expectation
value for this cluster: Aigrain et al. (2007) expected that there are
in total 11.3 transiting planets in NGC 2362 -- however, only 0.0 
of them were expected to be detectable by their survey.
This non-detection is for planets with 1.5 Jupiter radii 
and periods between 1 and 10 days (Miller et al. 2008).
With the $\sim 50$ mmag rms precision down to R=16.5 mag
(5 mmag rms down to R=14.5 mag, both for Jena), we can detect (young large) planets
down to $\sim 1$ Jupiter radius, given the typical radii and
brightness of the YETI targets stars. Given the YETI approach to observe
a certain field continuously for several days (24h per day),
e.g. continuously for up to $\sim 2$ weeks (e.g. the Tr-37 run 2010 Aug 26 to Sept 13), 
we would be able to detect all planets down to $\sim 1$ Jupiter radius with periods 
up to $\sim 14$ days. Hence, even if we would not detect any planets,
we would be able to place limits (for the frequency of young planets)
lower than the current limits for either the solar neighbourhood
or the NGC 2362 cluster (Miller et al. 2008).

To justify the YETI project, it is important that we understand 
why no transiting planets have been reported by the MONITOR survey so far.
It is possible
that this is due to the fact that the telescopes used in the MONITOR
project do not cover all longitudes on Earth (being located in Europe and
America only), so that continuous coverage is not possible.
Hence, transits can be missed; the light curves of stars from the
MONITOR project as published in the papers listed above also show
that the phase coverage is not complete.

\section{YETI target selection criteria and transit planet detection expectations}

For successful detection of a planetary transit it is very important to choose
regions on the sky where there is a high probability to observe this event. 

An interesting environment is represented by stars in a cluster. 
Open cluster surveys should help to clarify the factors that control the
formation and survival of planets by characterizing hot Jupiter populations
as a function of age, metallicity and crowding. 
A significant fraction of stars in the solar vicinity form in clusters
(Lada \& Lada 2003).
Stars in clusters also have the advantage that their age and distance is known,
which is otherwise often difficult to obtain precisely for isolated stars.

Wide-field CCD cameras on 1-2~m telescopes with $\sim 1^\circ$ Field-of-View (FoV)
are suitable for surveying stars in open clusters fields. 

An estimate of the number of expected transiting planets, $N_{\rm p}$, can 
be parameterized as

\begin{equation}
N_{\rm p} = N_{*} \cdot f_{\rm p} \cdot \rho _{\rm t} \cdot \rho _{\rm eff}.
\end{equation}

Here $N_{*}$ is the number of stars included in the transit search, 
$f_{\rm p}$ is the fraction of stars with close-in planets within 0.1 AU ($\sim 0.012$, Butler et al. 2006), 
$\rho _{\rm t}$ is the probability to view the orbit nearly edge-on ($\sim 0.1$
for close-in planets), 
and $\rho _{\rm eff}$ is a measure of the efficiency of the observation. 

The number of stars $N_{*}$ depends on the FoV of the telescope, 
the magnitude limits (excluding the brightest earliest stars and the
faintest latest stars), and the photometric precision. 
In wide-field transit searches or deep searches with large telescopes, 
usually thousands of stars are covered in the FoV. 
However, only the number of stars with sufficiently high S/N for
detecting transits are relevant in Eq. 1. For typical giant planets
around solar-type stars, this means the S/N must provide a detection limit
of $\sim 0.5\%$ dimming of the stellar light, or $\sim 5$ milli-mag (mmag) rms
for a significant detection. 
For young 10 Jup mass planets that are still contracting
(hence, large), the transit depth can be as large as
$\sim 80$ mmag (Burrows et al. 1997; Baraffe et al. 1998, 2000, 2001, 2002, 2003, 2008), 
see below for details.
Target stars must not be too faint, so that high-resolution spectroscopic
follow-up for determining the mass of the transiting companion from radial
velocities is possible with current-generation telescopes (8 to 10 meter mirrors),
i.e. down to about 16.5 mag (see e.g. the OGLE-TR-56 transit planet, Konacki et al. 2003).
The first observations 
with the Jena 90/60-cm telescope (90-cm mirror reduced
to 60-cm effective mirror diameter in Schmidt mode) in 2009
show that we can reach sufficient 
photometric precision down to R=16.5 mag
and 5 mmag rms down to R=14.5 mag
(Fig. 5). The limiting sensitivity of other telescopes for 
the Tr-37 cluster campaigns, from 0.4 to 2.6m, are similar within $\pm 1$ mag.
Hence, the number of stars $N_{*}$
in Eq. 1 in the FoV should be the number of stars brighter than R=16.5 mag
(the number of stars being too bright so that they saturate the CCD are negligible).

The parameter $\rho _{\rm eff}$ in Eq. 1 depends on many observational factors, 
e.g. weather, coordinates of targets and observatories; in Rauer et al. (2004),
who have observed with only one telescope (the Berlin Exoplanet Search telescope, BEST) 
at Tautenburg near Jena for 3 years, this factor was 0.7 - limited by weather
and the fact that only one telescope was used;
this factor is a period-dependent variable only calculable after the fact
as done for the BEST survey at Tautenburg by Rauer et al. (2004).
We can assume a larger number for the YETI survey, because we will use several telescopes
at many different longitudes on Earth to observe continuously. 
Hence, we can use $\rho _{\rm eff} \simeq 1$ for our calculations
(the factor should definitely lie between 0.7 and 1.0; if it would be 0.7
only, the number of expected observable transits listed in Table 1 would
decrease by only $30~\%$,
$\sim 1.0$ is a good approximation for planet periods up to $\sim 14$ days, 
detectable by us completly down to $\sim 1$ Jupiter radius given the
continuous observations.

See Table 1 for the clusters being observed so far 
and the expected number of transiting planets in the whole FoV
(including old field stars and among young members in the YETI clusters).
For young planets around young stars, we may assume a larger fraction
of transiting planets than for old planets around old stars:
If planets form partly by contraction, they are larger than old planets
(Burrows et al. 1997; Baraffe et al. 1998, 2000, 2001, 2002, 2003, 2008).
E.g., $\sim 3$ Myr young planets with 1 to 10 Jup masses are
expected to have radii of 1.4 times the radius of Jupiter, 
hence are a factor of 2 larger than old Gyr old 1 to 10 Jup mass 
planets (Burrows et al. 1997).
For planets with 1 to 12 Jup masses (radii from Baraffe et al. 2003, COND models)
transiting stars with 1 or 0.5~M$_{\odot}$ (radii from Baraffe et al. 1998, 2002),
the transit depth will have a maximum 
of $\sim 15$ to 80 mmag at $\sim 50$ to $\sim 100$ Myr,
respectively; for 40 Jup mass brown dwarfs (radii from Baraffe et al. 2003) 
eclipsing such stars, the transit depth will have a maximum 
of $\sim 60$ to 100 mmag at $\sim 5$ Myr, respectively; see Fig. 2.
For planets with 10 to 13 Jup masses (radii from Baraffe et al. 2000, 2001, DUSTY models)
transiting stars with 1 or 0.5~M$_{\odot}$ (radii from Baraffe et al. 1998, 2002),
the transit depth will have a maximum 
of $\sim 20$ to 80 mmag at $\sim 50$ to $\sim 200$ Myr, respectively; 
for 40 Jup mass brown dwarfs (radii from Baraffe et al. 2000, 2001)
eclipsing such stars, the transit depth will have a maximum 
of $\sim 70$ to 100 mmag at $\sim 10$ to 20 Myr, respectively.
For planets with 1 to 10 Jup masses (radii from Baraffe et al. 2008, both 
radiated and non-irradiated models)
transiting stars with 1 or 0.5~M$_{\odot}$ (radii from Baraffe et al. 1998, 2002),
the transit depth will have a maximum of $\sim 20$ to $70$ mmag at $\sim 50$ to 200 Myr,
respectively.
Even though these models may still be uncertain, in particular for
young ages and low masses, they tend to show that the transit depth
reaches a maximum at $\sim 10$ to 200 Myr (for 40 to 1 Jup mass companions
and 1 to 0.5~M$_{\odot}$ stars), see Fig. 2. After that maximum, the transit depth
decreases only slowly with age.
If the models are correct, we can use for young planets a factor of roughly 2
larger than for old planets (for number of expected detectable planets).

Through its first four months of operation, 
NASA's Kepler mission (Koch et al. 2010) detected 83 transit candidates with 
transit depths of at least 5 mmag and periods shorter than 10 days (Borucki et al. 2011). 
Given the Kepler target list of $\sim 160,000$ stars (Koch et al. 2010), this 
is an occurrence rate of 0.00052 short period transiting planets of this size per star. 
If we can assume that the rate of planet occurrence is the same for the Kepler target list 
and the YETI Tr-37 sample, then we can estimate the rate of planets detectable in Tr-37:
The Kepler target list has been carefully selected using the Kepler Input Catalog, 
and the Kepler targets are nearly all on or near the MS; the Kepler target list is dominated 
by F and G dwarfs, similar to the sample of Tr-37 members; the Tr-37 members
are of course younger (pre-MS) and, hence, larger than the Kepler MS targets;
the Kepler target list is, however, probably not a good match to a magnitude-limited sample 
in the Tr-37 FoV (field stars), which will have heavy contamination by giants;
however, we can apply the Kepler planet rate to the Tr-37 members.
Nevertheless, if we apply this estimate to Tr-37 (6762 stars in the Jena FoV, including 469 young members), 
and also correct for the size 
of young stars with young planets compared to old systems in the Kepler field,
we expect $\sim 0.5$ short-period planets 
among the known cluster members (and $\sim 3.8$ short-period planets total for the Tr-37 FoV). 
If we include periods out to 30 days, the number of candidates with transits deeper than 5 mmag 
detected by Kepler is 131, corresponding to $\sim 0.8$ expected planets among the known 
Tr-37 cluster members (and 5.9 total in the Tr-37 FoV). 

This estimate ($\sim 0.8$ transiting planets) is consistent with our other estimate 
(derived differently) given in Table 1. 
The estimates of expected detectable planets in Table 1 are upper limits in the sense,
that we may still miss some planets, if we always would have problems with weather and/or
technical issues, i.e. would never reach a truely contiunous monitoring, 
or due to the activity and, hence, intrinsic variability of the young targets
(however, $\sim 10$ to 100 Myr old stars like weak-line T Tauri stars and zero-age
main sequence stars already show much lower variability amplitudes compared
to $\sim 1$ Myr young classical T Tauri stars).
The estimates for planets in Table 1 are also lower limits, 
because the number of young members in the clusters known 
(and given in Table 1) are also lower limits; in the YETI survey, we will find new members
by photometric variability and follow-up spectroscopy.
Using the monitoring campaigns with three one- to two-week runs in three consecutive months, 
repeated in two or three consecutive years, it will be possible to detect most of 
the transiting planets with periods up to $\sim 30$ days.

We show in Fig. 6 below that we are indeed able to combine the photometric
data points from different telescopes (with different mirror sizes,
different detectors, and different weather conditions) with sufficient
precision to detect transits: Fig. 6 shows a preliminary light curve
for an eclipsing double-lined spectroscopic binary in the Tr-37 field;
this is probably not a young member of Tr-37, but still to be confirmed.

Even though transiting planets have not yet been found in clusters
(most of the clusters surveyed for transits are several Gyr old),
it is not yet known, whether the occurence rate of (transiting) planets is
lower in clusters compared to non-cluster field stars. If hot (transiting) Jupiters
do not survive as long as the cluster age, then young clusters may still
have more planets than old clusters. On the other hand, stars in young clusters
are more active than in old clusters, which makes it more difficult to
detect and confirm young transiting planets. It is also not yet known how long planet
formation (and migration) lasts, i.e. whether planet formation is possible
during the (small) age of the YETI clusters. In this project, we aim to clarify
those questions.

We argue that the fraction of transiting planets expected and found
in the MONITOR survey is lower than expected in the YETI survey, 
because we obtain continuous monitoring
with telecopes covering all longitudes on Earth, i.e. 24h per day for several
consecutive days. With such continuous observations, we will be able to
detect any transit with a depth of at least about 5 mmag
rms (down to 14.5 mag), the mean YETI sensitivity, 
for every planet with an orbital period lower than the survey -
or, in order to detect the periodicity of the candidate, lower than about
half the survey duration (somewhere between a few days and a few months).

The YETI cluster target selection criteria to study planet formation by transit observations 
are therefore as follows:
\begin{itemize}
\item Young age: At least one Myr, so that planets may already exist,
and younger than $\sim 100$ Myr, the PMS time-scale of the lowest mass stars.
\item Intermediate distance: Neither too close (otherwise the field with enough stars
would be too large on the sky) nor too distant (otherwise the stars would be too faint),
roughly 50 to 1000 pc.
\item Size of the cluster on sky: Roughly $1^{\circ} \times 1^{\circ}$,
i.e. well suited to the typical field sizes set be the YETI telescope optics and CCD sizes;
in case of smaller FoVs and mosaicing, the cadence should still be high
enough to obtain sufficient data points during the typical transit duration.
\item Clusters not studied before with similar or better observations,
e.g. like NGC 2264 with CoRoT.
\item As many young stars as possible in a useful magnitude range in VRI:
Not too bright (fainter than $\sim 10$ mag) to avoid saturation 
and not too faint (brighter than 16.5 mag) not to lose sensitivity
(even if different exposure times are used for stars of different 
brightness) 
and to be able to do RV follow-up spectroscopy of transit candidates 
(the faintest host star where this was successfully done is OGLE-TR-56 with V=16.6 mag, Konacki et al. 2003).
\item Location on sky, so that many telescopes can observe and monitor it continuously.
\end{itemize}

The number of confirmed transit planet detections is 134 (e.g. exoplanet.eu),
relatively small when compared with early predictions (Horne 2003).

The origin of this discrepancy may have several causes: \\
(i) simplifying assumptions in the noise properties that govern the detection limits in
previous predictions - in many cases red instead of white noise may be
dominating (Pont et al. 2006), \\
(ii) unaccounted errors in aperture photometry on non-auto- guided telescopes, \\
(iii) overestimates of the fraction of stars that are suitable 
as targets for transit surveys, and \\
(iv) underestimates of the need for continuous observations. 

Another topic whose consequences have
only been recognized during the course of the first transit searches is the
problem of false alarms caused by other stellar combinations that may
produce transit-like light curves. Of these, the most notorious case may be
that of an eclipsing binary star located within the point spread
function (PSF) of a brighter star (Brown 2003). An increasing number of
tools are now available to recognize false alarms: (i) precise analysis of
transit shapes or durations, (ii) color signatures, (iii) variations in the
positioning of the stellar PSF. The effectiveness of several of these tools
depends strongly on the information (e.g., temperature, radius) that is
available for the host star, and underlines the need for auxiliary
observations in transit detection experiments.

In practice, it is not always possible from the light curve alone to
exclude false positives, such as background eclipsing binaries blended
with the target.
We obtain follow-up spectroscopy to aid in examining these
possibilities, as well as to measure the stellar properties temperature,
rotational velocity, gravity, and metallicity,
and to measure the mass of the companion.
Except when TTV signals are observed,
one always needs follow-up spectroscopy to determine the mass of the companion.

\begin{table*}
\begin{tabular}{lcccccccc} 
\multicolumn{9}{c}{{\bf Table 1. First two target clusters}} \\ \hline
Cluster & \multicolumn{2}{c}{central coordinates} & age    & distance & \multicolumn{3}{c}{\scriptsize number of stars with R$\le 16.50$ mag} & {\scriptsize expected no. of transit planets} \\
name    & RA J2000.0 & Dec J2000.0 & [Myr] & [pc]     & {\scriptsize members} & {\scriptsize members in}  & {\scriptsize stars in} & {\scriptsize in the Jena FoV (a)} \\ 
        & [hh:mm:ss] & [dd:'':""] &        &          & {\scriptsize in total}    & {\scriptsize Jena FoV} & {\scriptsize Jena FoV (b)}  & {\scriptsize (and among members)}  \\ \hline
Trumpler 37 & 21:38:09 & +57:26:48 & 7     & 870      & 614        & $\ge 469$     & 6762     & $\sim 8.1$ ($\sim 1.1$) \\ 
25 Ori      & 05:24:45 & +01:50:47 & 7-10  & 323      & 179        & $\ge 108$     & 1045     & $\sim 1.3$ ($\sim 0.3$) \\ \hline 
\end{tabular}

Remarks: (a) According to Eq. 1 and Sect. 3. (b) Jena FoV is $53^{\prime} \times 53^{\prime}$ for STK, see Table 2.

References for coordinates, age, distance, and members: 25 Ori: Kharchenko et al. (2005), Brice\~no et al. (2005, 2007);
Trumpler 37: See Sect. 6.
\end{table*}

\section{The YETI telescope network for continuous monitoring}

Given the target selection criteria listed above, we have selected the clusters 
Tr-37 and 25 Ori (Table 1) as the first two clusters to be observed, 
several more will follow later in order
to cover the full age range to study formation and early evolution of planets.
Additional clusters to be observed fully or partly in future years are
Pleiades, NGC 2244, $\alpha$ Per, h and $\chi$ Per, Collinder 69, $\sigma$ Ori, IC 2602, etc.
If we will observe approximately ten clusters in the project,
each cluster for 2 or 3 years, and not more than two different clusters per year,
then the whole YETI project will last for more than ten years.
Most of the observatories participating can allocate their telescopes for
all nights in all runs for the whole duration of the YETI project
(mostly already guaranteed, partly by internal proposals), while for some of
the participating telescopes, we have to apply for time (e.g. Sierra Nevada,
Calar Alto, Mauna Kea).

The clusters Tr-37 (Fig. 1) and 25 Ori have been observed since 2009.
The cluster Tr-37 was observed by most
of the participating telescopes in 2010 during the three runs
August 3/4 to 12/13, August 26/27 to September 12/13, and September 24/25 to Sept 30/Oct 1,
i.e. 35 nights in total. 
The cluster 25 Ori was observed simultaneously and continuously by most
of the participating telescopes
2010 December 10/11 to 17/18, 2011 January 13/14 to 23/24, and Feb 16/17 to 27/28.
Additional nights for the 2011 and subsequent observing seasons are
expected to be allocated on all the YETI telescopes, 34 nights for Tr-37 in 2011
and 29 nights for 25 Ori in the (northern) winter 2011/12.
We also observed a field at the edge of the Pleiades cluster
(Eisenbeiss et al. 2009, Moualla 2011). First results from Tr-37 are reported below. 

\begin{table*}
\begin{tabular}{lllllccc} 
\multicolumn{8}{c}{{\bf Table 2. Telescope network (1)} (sorted by longitude)} \\ \hline
Observatory         & Long.     & Lat.    & Mirror dia- & CCD type        & no. of    & size of field & Ref. \\ 
                    & [deg]     & [deg]   & meter [m]   & (camera)        & pixels    & [min x min] & \\ \hline
Gunma/Japan         & 139.0 E  & 36.6 N  & 1.50        & Andor DW432              & 1250 x 1152 & 12.5 x 12.5 & (2) \\ 
Lulin/Taiwan        & 120.5 E  & 23.3 N  & 1.00        & Marconi CCD36-40 PI1300B & 1340 x 1300 & 22 x 22 & \\
                    &          &         & 0.41        & E2V 42-40 (U42)          & 2048 x 2048 & 28 x 28 & \\
Xinglong/China      & 117.6 E  & 40.4 N  & 0.90 (3)    & E2V CCD203-82            & 4096 x 4096 & 94 x 94 & Wu07 \\
Nainital/India      & 79.5 E   & 29.4 N  & 1.04        & TK2048E                  & 2000 x 2000 & 13 x 13 & \\
Byurakan/Armenia    & 44.3 E   & 40.3 N  & 2.60        & SCORPIO Loral            & 2058 x 2063 & 14 x 14 &      \\
Rozhen/Bulgaria     & 24.7 E   & 41.7 N  & 0.60        & FLI ProLine 09000        & 3056 x 3056 & 17 x 17 & (4) \\
                    &          &         & 0.60 (5)    & FLI ProLine 09000        & 3056 x 3056 & 27 x 27 & (4) \\
                    &          &         & 0.70 (6)    & FLI ProLine 16803        & 4096 x 4096 & 73 x 73 & (4) \\
                    &          &         & 2.00        & Princ. Instr. VersArray:1300B & 1340 x 1300 & 6 x 6 & (4) \\
Star\'a Lesn\'a     & 20.3 E   & 49.2 N  & 0.50        & SBIG ST10 MXE            & 2184 x 1472 & 20 x 14 & \\
(Slovak Rep.)       &          &         & 0.60        & SITe TK1024              & 1024 x 1024 & 11 x 11 & \\
                    &          &         & 0.25        & SBIG ST10 MXE            & 2184 x 1472 & 43 x 29 & \\
Toru\'n/Poland      & 18.6 E   & 53.1 N  & 0.90 (3)    & SBIG STL-11000           & 4008 x 2672 & 48 x 72 & \\
Jena/Germany        & 11.5 E   & 50.9 N  & 0.90 (3)    & E2V CCD42-10 (STK)    & 2048 x 2048 & 53 x 53 & Mug10 \\
                    &          &         & 0.25 (7)    & SITe TK1024  (CTK)    & 1024 x 1024 & 38 x 38 & Mug09 \\
                    &          &         & 0.25 (8)    & E2V CCD47-10 (CTK-II) & 1056 x 1027 & 21 x 20 & Mug11a \\  
                    &          &         & 0.20       & Kodak KAF-0402ME (RTK) & 765 x 510   & 8 x 5   & Mug11b \\
S. Nevada/Spain & 3.4 W & 37.1 N & 1.50  & EEV VersArray:2048B   & 2048 x 2048 & 8 x 8   &     \\
Calar Alto/Spain    & 2.5 W    & 37.2 N  & 2.20 (9)    & SITe1d (CAFOS)        & 2048 x 2048 & 16 x 16 & (10) \\
Armazones/Chile     & 70.2 W   & 24.6 S  & 0.15        & Apogee U16M KAF-16803 & 4096 x 4096 & 162 x 162 & (11) \\
CIDA/Venezuela      & 70.9 W   & 8.8 N   & 1.00        & Quest-I CCD Mosaic    & 8000 x 8000 & 138 x 138 & Ba02  \\
                    &          &         & 1.00 (5)    & FLI Proline E2V42-40  & 2048 x 2048 & 19 x 19 & \\
Stony Brook/USA     & 73.1 W   & 40.9 N  & 0.37       & SBIG ST1001E KAF-1001E & 1024 x 1024 & 17 x 17 &    \\
Swarthmore/USA      & 75.4 W   & 39.9 N  & 0.62        & Apogee U16M KAF-16803 & 4096 x 4096 & 26 x 26 &      \\
Gettysburg/USA      & 77.2 W   & 39.8 N  & 0.40        & SITe 003B             & 1024 x 1024 & 18 x 18 & (12) \\
Tenagra II/USA      & 110.5 W  & 31.3 N  & 0.81        & SITe SI003 AP8p       & 1024 x 1024 & 15 x 15 &    \\ 
Mauna Kea/Hawaii    & 155.5 W  & 19.8 N  & 2.20 & 8 CCD chips for mosaic & {\scriptsize 8 x 2048 x 4096} & 33 x 33 & \\ \hline
\end{tabular}

Remarks: (1) Listed are only those, from which we have obtained (or proposed) photometric monitoring 
data so far; 
(2) www.astron.pref.gunma.jp/e/inst$\_$ldsi.html;
(3) 0.60m in Schmidt mode;
(4) www.nao-rozhen.org/telescopes.fr$\_$en.htm;
(5) with focal reducer;
(6) 0.50m in Schmidt mode;
(7) until July 2010;
(8) since August 2010;
(9) by open time observing proposals;
(10) w3.caha.es/CAHA/Instruments/CAFOS/cafos$\_$overview.html;
(11) www.astro.ruhr-uni-bochum.de/astro/oca/vysos6.html;
(12) www3.gettysburg.edu/$\sim$marschal/clea/obshome.html. \\
Ref.: Mug09 - Mugrauer 2009; Mug10 - Mugrauer \& Berthold 2010; Mug11a,b - Mugrauer 2011a,b (in prep);
Wu07 - Wu et al. 2007; Ba02 - Baltay et al. 2002.
\end{table*}

Given the problems listed in the previous section
and in order not to miss a transit and to determine the periodicity in the
transit-like variability, it is important to monitor the targets (almost) continuously,
i.e. 24 hours per day for the whole run of at least several days.
Therefore, we established a network of several observatories around the
world covering many longitudes, to that we can observe the YETI targets continuously.

The participating observatories are listed in Table 2 with their telescopes and instruments.

\section{Additional science projects}

Although the main thrust of this project is the search for young
transiting planets, the observations we are collecting also lend
themselves to studies of variability phenomena on different
time-scales, and can be co-added as well to provide very deep imaging
of the YETI target clusters. An additional component of the project is a
theoretical study of the interiors of young transiting planets to
provide a framework for interpreting the YETI discoveries. We describe
these topics below.

\subsection{Other variability phenomena}

The precise and nearly continuous time series photometry gathered for
this project enables us to investigate rotational periods for cluster
members and field stars (see e.g. Fig. 3; first results on Tr-37 data from 
Jena from 2009 can be found in Berndt et al. 2011), longer-term cycles
(Fig. 4), and irregular variability like flares.
We can also detect eclipsing binaries comprised of two stars, 
or two brown dwarfs, or a brown dwarf orbiting a star.
From eclipsing binaries among young cluster members, we can derive
mass-luminosity relations, which are one of the 
main uncertanities in obtaining luminosity and mass functions
and of pre-main sequence populations as well as ages and masses
of individual stars.
We plan to use multi-color photometry
to refine age, distance, and extinction measurements
for the clusters, e.g. from the color-magnitude or color-color diagrams.
 
\subsection{Deep imaging}

We can add up many images obtained per cluster to get much deeper photometry.
We have already tested this using a field at the edge of the 
Pleiades, where we could find several brown dwarf candidates by
RIJHK colors (Eisenbeiss et al. 2009), for which we have
recently obtained follow-up optical and infrared spectra with the ESO VLT
(Seeliger 2011).
We also applied this technique very successfully by combining co-added 
optical I-band data of $\sim 180$ square degrees across the Orion OB1 region, 
with 2MASS JHK data (Downes et al. 2008; Downes et al., in preparation), 
to identify several thousand young very low-mass stars and brown dwarfs, 
for many of which we have confirmed membership through follow-up spectroscopy
(Downes et al., in preparation).

Co-added individual frames from the larger telescopes will also provide 
a more sparsely sampled, but much deeper time series, which will allow us 
to carry out systematic variability studies amongst the young brown dwarfs of 
the YETI target clusters.

Similar studies can be done for all clusters observed by YETI.
Such a study can reveal the very low-mass population including massive
brown dwarfs, i.e. the low-mass end of the mass function.
We can re-determine cluster distance, age, and mass function
based on a larger set of photometric and spectroscopic data
obtained for a larger sample of cluster members.
Given the large FoV, we can probe possible variations of
the mass function and brown dwarf density with the distance to the center 
of the cluster. 

\begin{figure*}
{\includegraphics[angle=0,width=16cm]{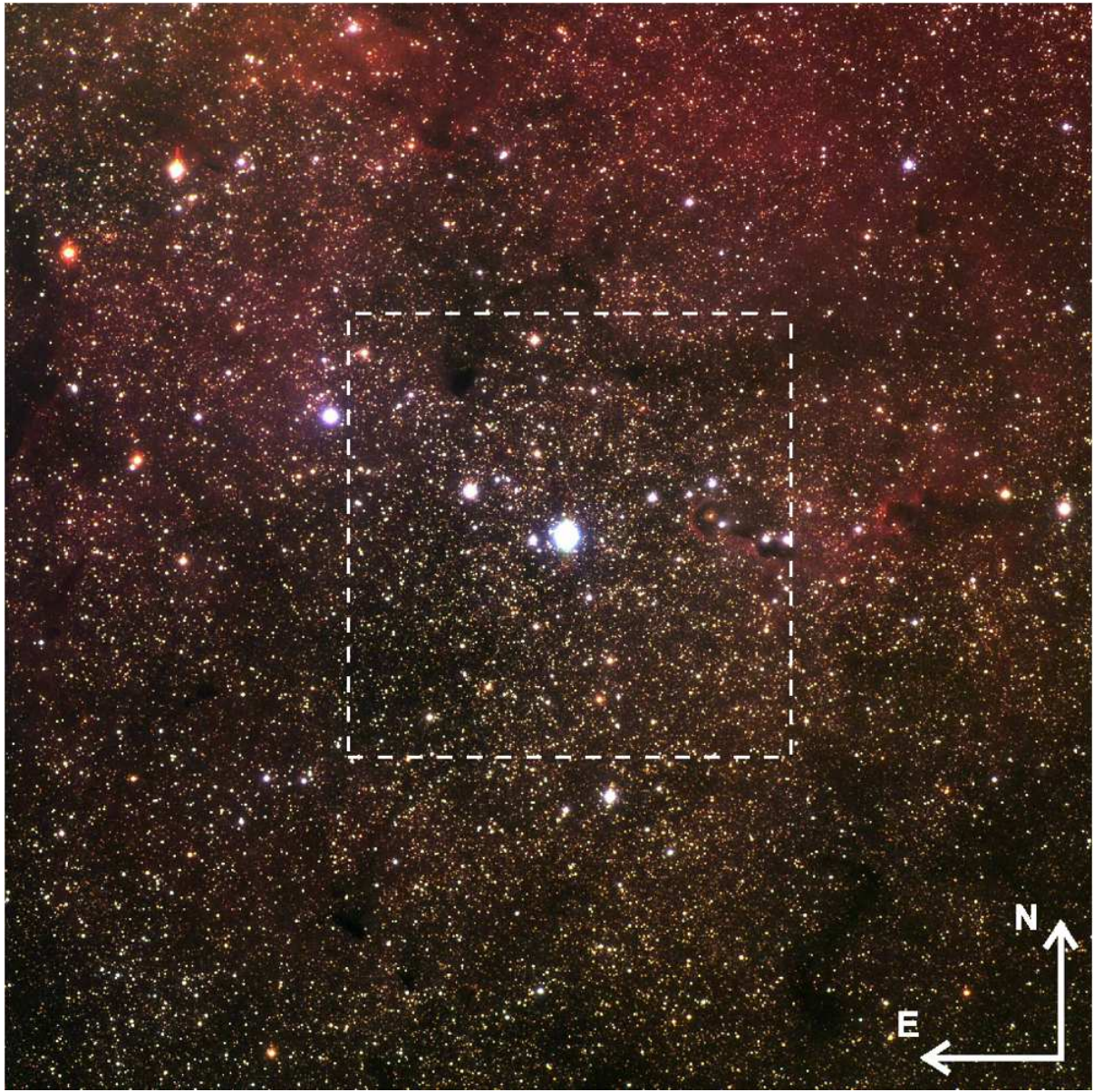}}
\caption{A BVR three-color composite image of the Tr-37 cluster as observed
with STK at University Observatory Jena in July 2009. The image
is a mosaic of nine STK images, each the composite of three 60s integrations 
taken in the B-, V-, and R-band. The total FoV shown is $2.1^{\circ} \times 2.1^{\circ}$ 
(north is up, and east to the left). The central dashed box indicates the 
STK $53^{\prime} \times 53^{\prime}$ FoV monitored in Jena only in 2009
and with the YETI consortium in 2010.
}
\end{figure*}

Deep images can also be used to study the proper motion
compared to previous images, hence to detect new cluster members.
We note that for deep imaging, we do not plan to add up all
data from all different telescopes together into just one deep image,
because of differences in FoV size, pixel scales, seeing, etc.,
but we will use the data from the largest telescope(s) 
(and also the one with the largest FoV) to add up all images 
from that telescope(s) into one deep image (one per detector).

\begin{figure}
{\includegraphics[angle=0,width=8cm]{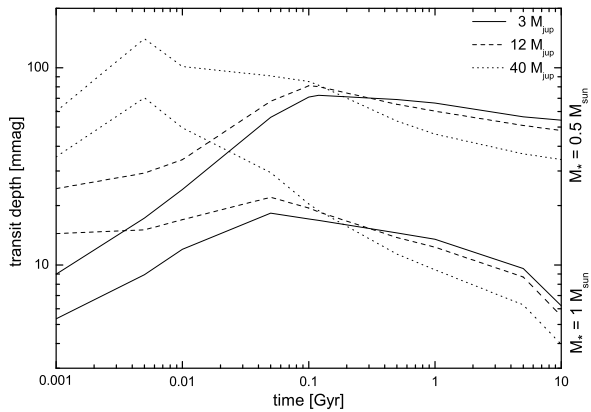}}
\caption{Expected transit depth (in milli-mag) 
versus ages (in Gyr) for 40 Jup mass brown dwarfs (dotted lines)
as well as 12 and 3 Jup mass planets (dashed and full lines, respectively)
transiting 0.5 (top) and 1~M$_{\odot}$ (bottom) stars according to 
Baraffe et al. (1998, 2002, 2003). These theoretical calculations show
that the transit depth reaches a maximum at 5 to 100 Myr due to
different contraction time-scales of stars and sub-stellar objects -
hence, our target cluster selection.}
\end{figure}

\subsection{Interior models for giant planets}

Widely accepted models for giant planets assume three layer structures 
composed of a central core of rock or ices and two fluid envelopes 
above (Guillot 1999 a,b). The location of the layer boundaries and the 
composition within the layers are subject of an optimization procedure 
with respect to observational constraints, e.g. gravitational moments 
for the solar giant planets and possibly the Love number (Kramm et al. 2011),
if known for an extrasolar giant planet. 

The main input in interior models is the equation of state (EoS) for H, 
He, H$_{2}$O, NH$_{3}$, CH$_{4}$, or rock material for the respective extreme conditions 
(up to several 10 megabars and several 10000 K). We perform ab initio 
molecular dynamics simulations which yield accurate EoS data and electrical 
conductivities for H (Holst et al. 2008, Lorenzen et al. 2010), He 
(Kietzmann et al. 2007), H-He mixtures (Lorenzen et al. 2009), and water 
(French et al. 2009, 2010). Besides the generation of wide range EoS data
tables, we study the high-pressure phase diagram, nonmetal-to-metal 
transitions, and demixing phenomena which are important for interior models. 

Based on these results, we have determined the structure of giant planets 
and calculated their cooling history: Jupiter (Nettelmann et al. 2008), 
the hot Neptune GJ436b (Nettelmann et al. 2010; Kramm et al. 2011), 
and for Uranus and Neptune (Fortney \& Nettelmann 2010; Redmer et al. 2011). 

Within the YETI project we plan to develop interior models especially 
for young planets that are detected within the observational campaigns. 
Then, we can study the formation and early evolution of giant planets.

\section{First results for Trumpler 37}

The open cluster Trumpler 37 (Tr-37), first studied by Trumpler (1930)
and Markarian (1952), is embedded in the HII region IC 1396,
and is the nucleus of the Cep OB2 association (Simonson 1968).
The V=4 mag M2Ia super-giant $\mu$~Cep is a probable member of the cluster,
but outside of the Jena FoV (and outside the mosaic in Fig. 1).
Comparison of the Tr-37 main sequence to that of Upper Scorpius yielded
a distance modulus of 9.9 to 10 mag, i.e. $\sim 1000$ pc (Garrison \& Kormendy 1976);
from the MS life-time of the earliest member, the O6e type star HD 206267 (V=5.6 mag),
they conclude an age of 2 to 4 Myr.
Marschall \& van Altena (1987) studied the proper motions of $\sim 1400$ stars
within $1.5^{\circ}$ of HD 206267 down to V=15 mag and found $\sim 500$ kinematic members.
Marschall et al. (1990) then determined the age of the cluster by the
MS contraction time of those members which just have reached
the zero-age main-sequence (ZAMS), to be $\sim 6.7$ Myr;
later, an age range from 3 to 10 Myr was considered
(Contreras et al. 2002; Sicilia-Aguilar et al. 2004b, 2005).
The current best age estimate is $\sim 4$ Myr (Kun, Kiss, Balog 2008),
and the latest distance estimate is $870 \pm 70$ pc (Contreras et al. 2002).

\begin{figure}
{\includegraphics[angle=270,width=8cm]{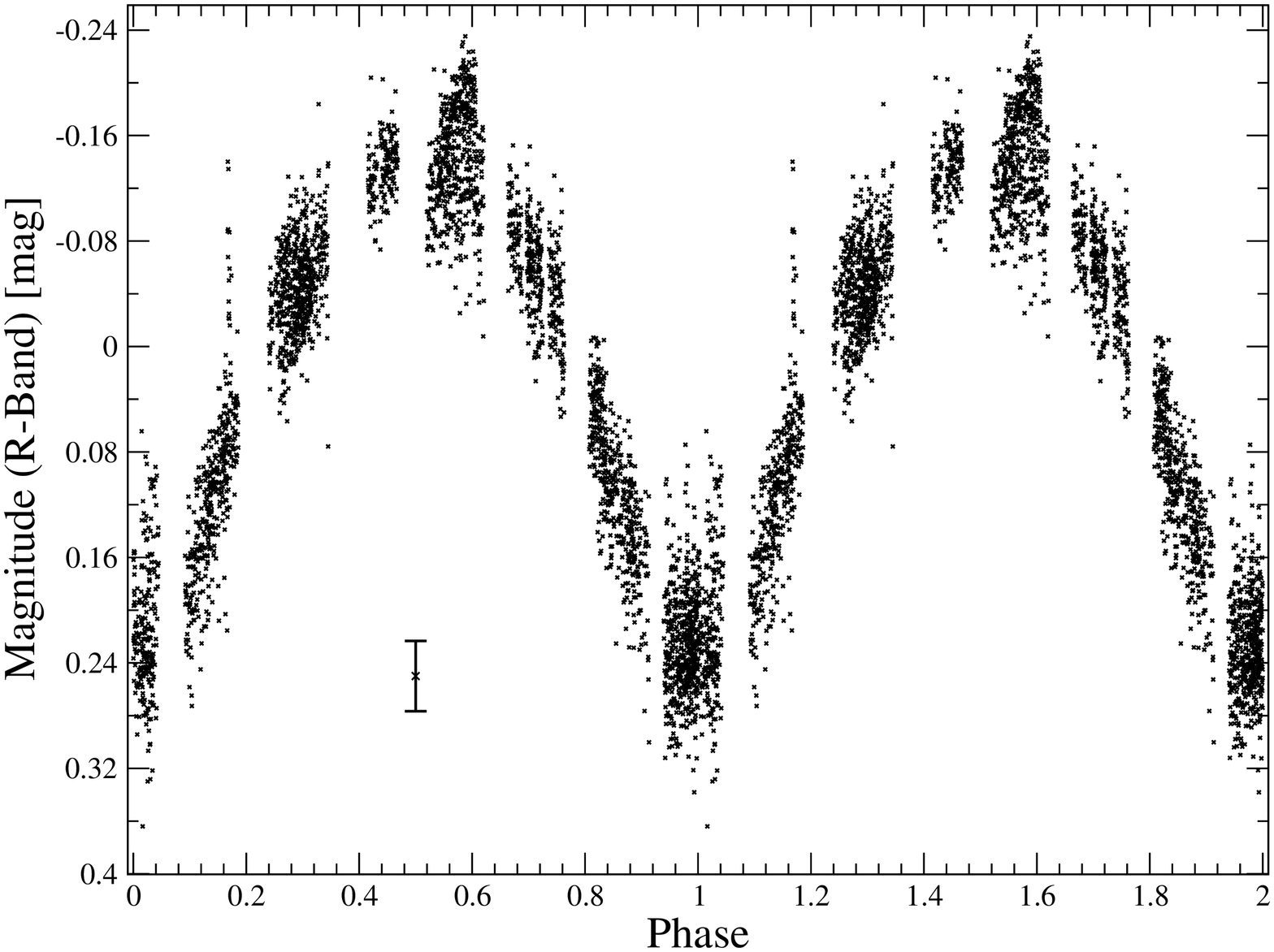}}
\caption{Phase-folded R-band light curve from Jena STK data from July to Nov 2009
for the star 2MASSJ21353021+5731164.
The error bar in the lower left is the typical (mean) photometric error.
The star is a classical T Tauri member of Tr-37 with spectral type K6 
according to optical spectra (Sicilia-Aguilar et al. 2005, 2006b).
We find R=15.8 mag and V=16.9 mag, consistent with K6,
and a rotation period of $\sim 3.5$ days,
it has a relatively large peak-to-peak amplitude of $\Delta R \simeq 0.5$ mag, which has been
observed also in a few other classical T Tauri stars before.}
\end{figure}

A three-color BVR composite image of the Tr-37 cluster obtained with the STK CCD camera
at the Jena 60-cm Schmidt telescope is shown in Fig. 1.

\begin{figure}
{\includegraphics[angle=270,width=8cm]{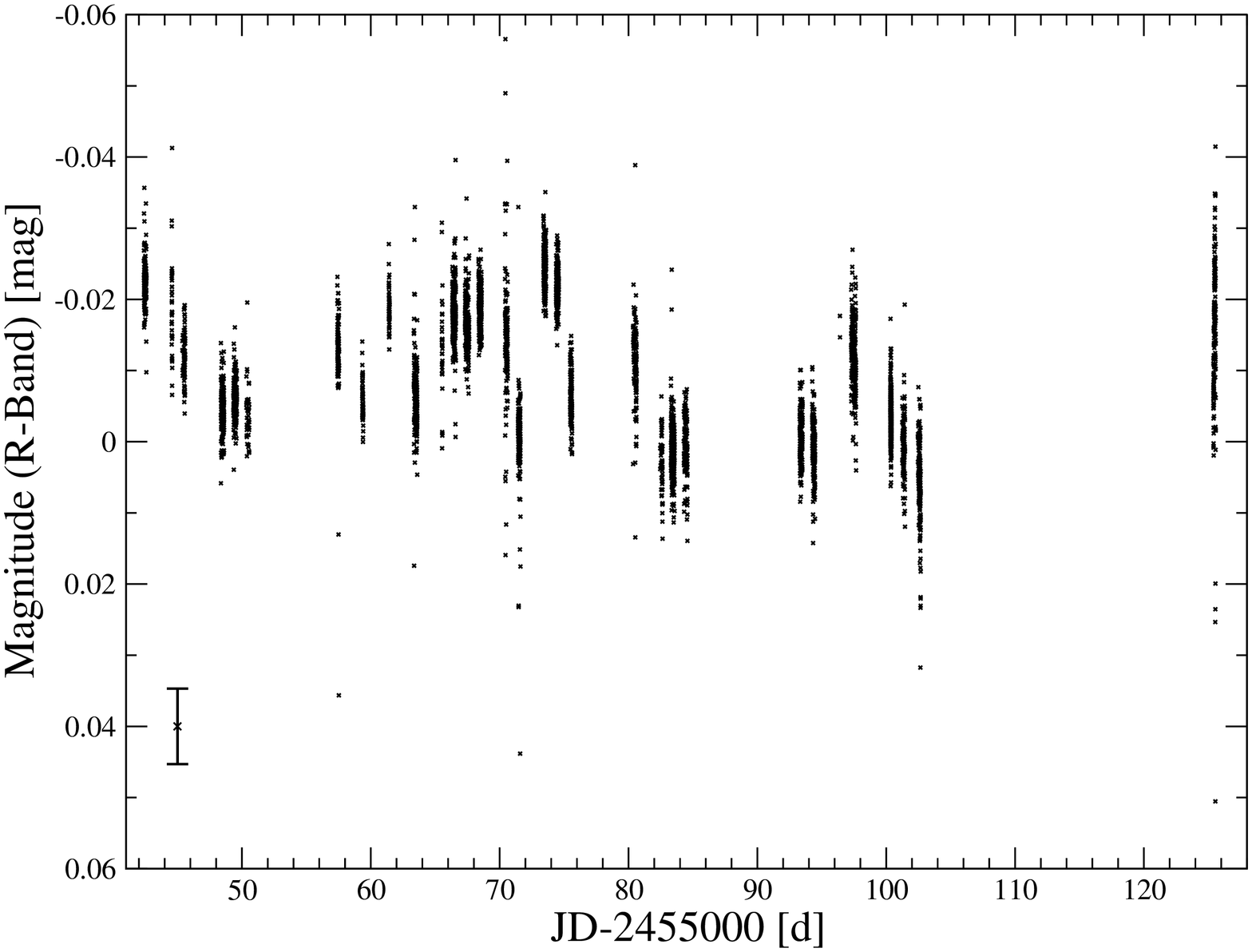}}
\caption{R-band light curve from Jena STK data from July to Oct 2009
for the star MVA 1312, spectral type B4 (Contreras et al. 2002, Sicilia-Aguilar et al. 2005),
it has a proper motion membership probability of 0.84 (Marschall \& van Altena 1987).
The observing date is given in JD since 2009 June 17.
The error bar in lower left is the typical (mean) photometric error.
We find R=10.6 mag and V=10.3 mag, consistent with early B.
The star shows variability on short (nightly) and longer time-scales (days to weeks),
but we also have strong gaps in the light curve from one observatory (Jena).
This shows the need for continuous monitoring.}
\end{figure}

\begin{figure}
{\includegraphics[angle=270,width=8cm]{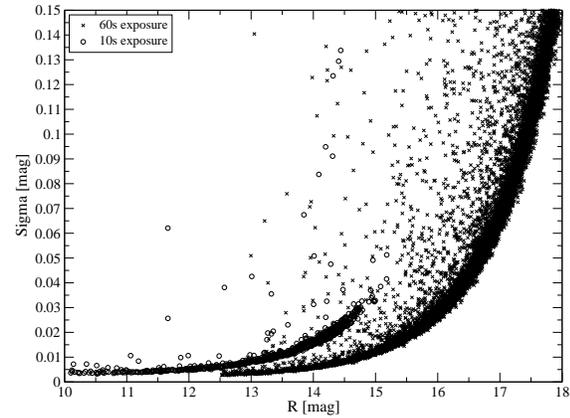}}
\caption{Photometric precision achieved (in mag) versus apparent photometric
brightness (in mag in R-band with Jena 90/60-cm telescope
in the night 2009 Sept 9/10) with 6762 stars in the Tr-37 field. 
A precision of better than 50 milli-mag rms
(sufficient to detect transits) is achieved for all stars brighter than 16.5 mag,
and a precision of 5 milli-mag rms for all stars brighter than R=14.5 mag.}
\end{figure}

\begin{figure}
{\includegraphics[angle=270,width=8cm]{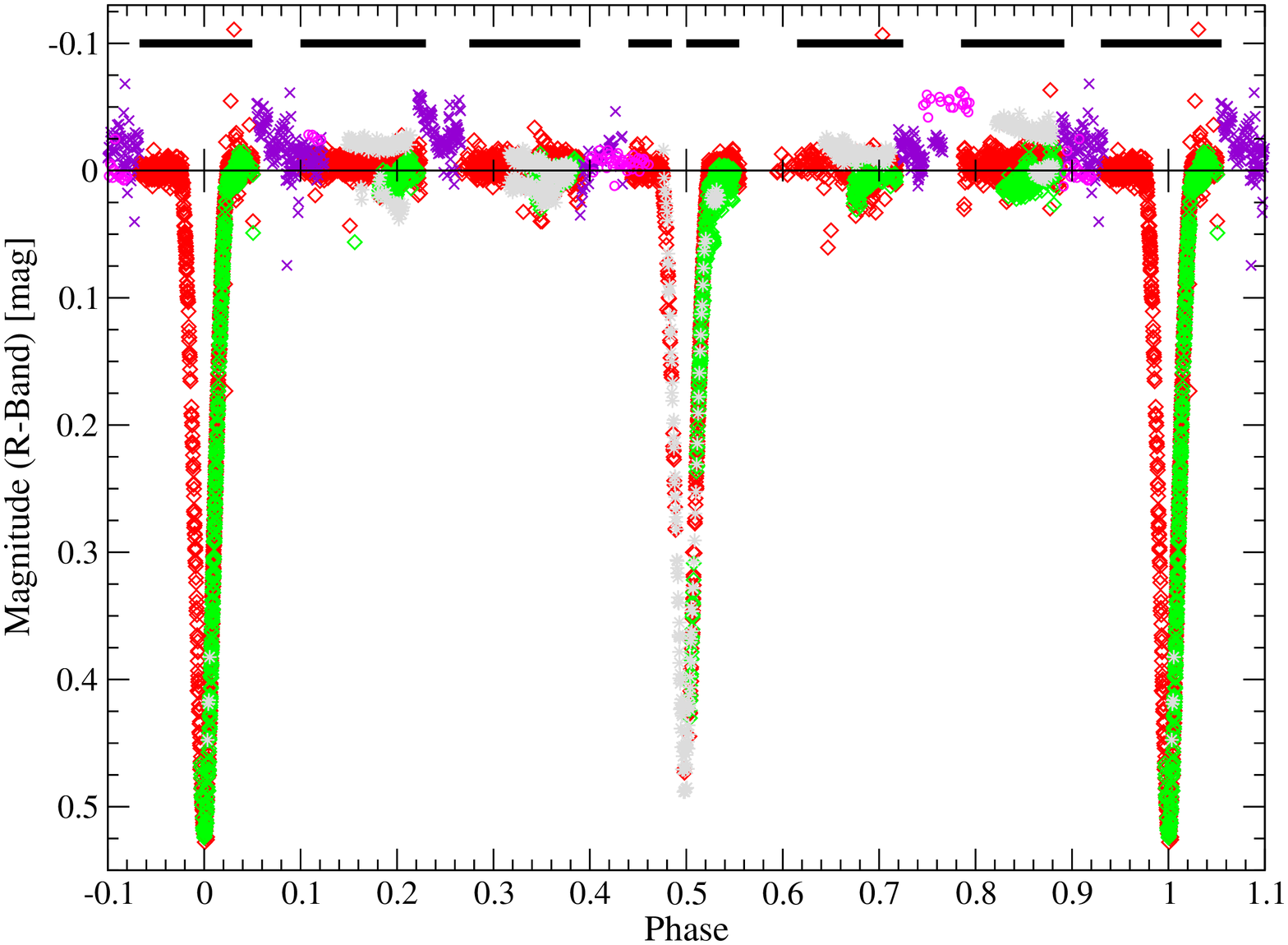}}
\caption{A preliminary light curve of an eclipsing star in the Tr-37 field
phased to a 6.005 day orbital period. Data from Jena
are shown as red and green diamonds (for 10s and 60s exposures, respectively). 
From Jena alone, we could not fully cover the secondary (shallower) eclipse (phase 0.5).
The bars on the figure top show the phases covered from Jena.
The other (colored) symbols show the data from four other telescopes:
Brown plusses from Byurakan/Armenia (60s), pink crosses from Xinglong/China (10s),
grey stars from Swarthmore/USA (60s), and pink circles from Lulin/Taiwan (10s).
This example shows how important
it is to cover all longitudes on Earth, hence the YETI telescope network.
Follow-up spectroscopy has shown that this star is a double-lined spectroscopic binary.
This particular star is probably not a young member of the Tr-37 cluster, 
but follow-up observations are still ongoing. Final results will be reported
later in Errmann et al. This figure shows that we can successfully combine
data from different telescopes with different CCDs and different ambient conditions.}
\end{figure}

A total of 732 members or member candidates of the cluster were found by H$\alpha$ emission,
ZAMS or pre-MS location, Lithium absorption, X-ray emission, infrared excess
emission, proper motion or radial velocity (Marschall \& van Altena 1987; 
Marschall et al. 1990; Contreras et al. 2002; Sicilia-Aguilar et al. 2004a,b;
Sicilia-Aguilar et al. 2005; Sicilia-Aguilar et al. 2006a,b; Mercer et al. 2009).
Some of the candidates may not be members, e.g. stars with either radial velocity or
proper motion consistent with membership, but no or very weak Lithium absorption,
while true members may still be missing, e.g. faint very low-mass members.
We are taking low- and high-resolution spectra of hundreds of stars in 
YETI $\sim 1^{\circ} \times 1^{\circ}$ FoV with MMT/Hectochelle,
in particular of suggested but uncertain member candidates, in order to confirm
or reject them as members based on radial velocity and Lithium absorption
(Errmann et al., in prep.).

\begin{figure*}
{\includegraphics[angle=0,width=16cm]{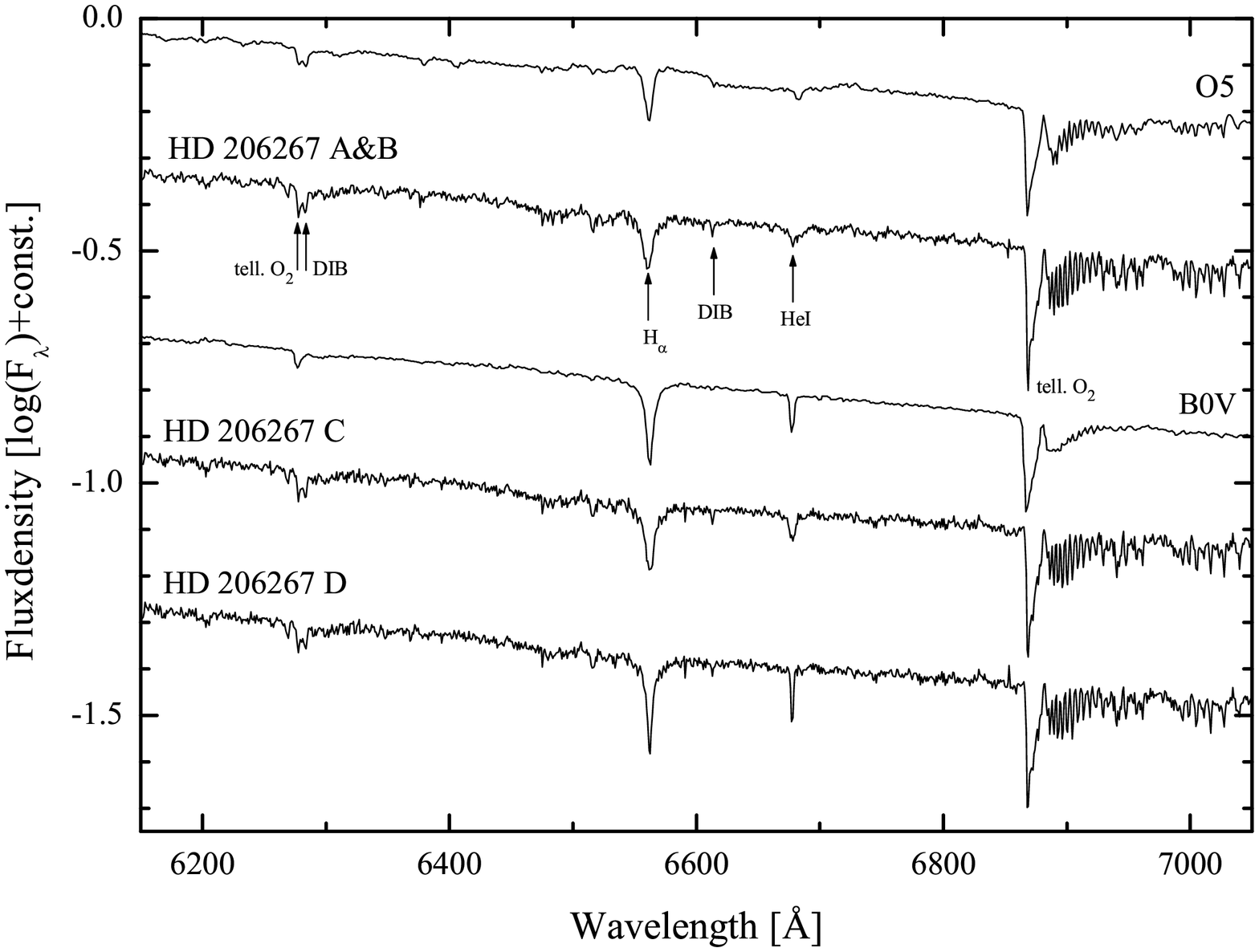}}
\caption{An optical spectra of HD 206267A+B (together), C, and D all obtained with the FIASCO spectrograph
at the Jena 90-cm telescope on 2010 June 7, shown together with template spectra
(O5 star HD 93250 and B0V star HD 36512 from Le Borgne et al. 2003). 
The spectrum of HD 206267A+B is consistent with a spectral type of O6 (as in Simbad and
Sota et al. 2011), 
and the spectra of both HD 206267 C and D are consistent with a spectral type of 
B0V (as in Simbad and Sota et al. 2011); HD 206267A+B as well as C and D show H$\alpha$ and He absorption 
as well as diffuse interstellar absorption bands and telluric oxygen.}
\end{figure*}

\begin{figure}
{\includegraphics[angle=270,width=8cm]{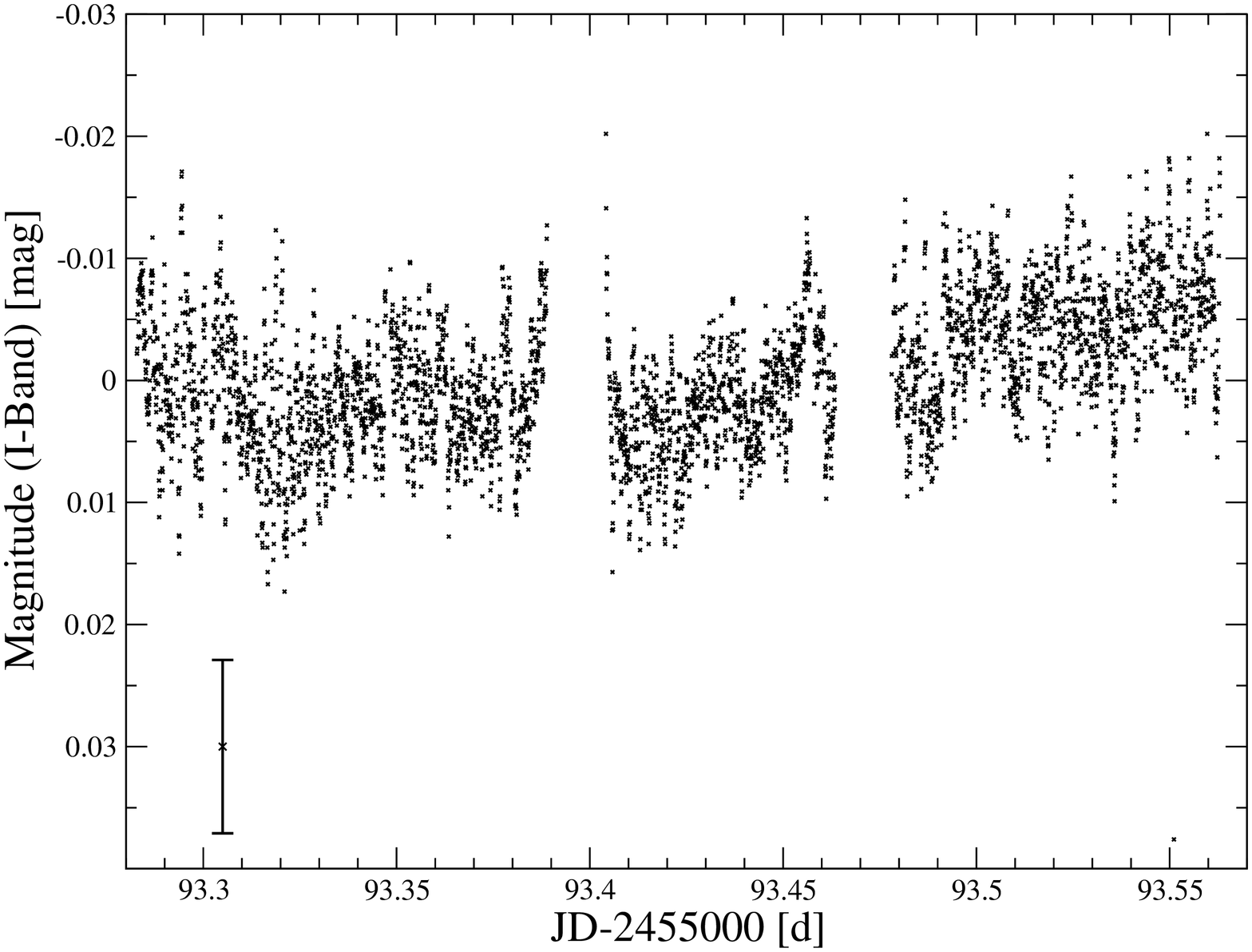}}
\caption{I-band light curve from Jena RTK data from 2009 Sept 18
for the star HD 206267A+B (spectroscopic binary), the brightest member of 
Tr-37 in the Jena FoV, I=5.6 mag, V=5.6 mag, and spectral type O6 (Simbad, Sota et al. 2011, 
and our Fig. 7).
The error bar in the lower left is the typical (mean) photometric error.
The membership probability of this star to Tr-37 is 0.67 (Marschall \& van Altena 1987).
The time resolution is 40s, because each data point plotted is a mean of
eight exposures of 5s.}
\end{figure}

We present here preliminary results 
including a few examplary light curves from the 2009 data,
obtained in Jena only with the 90/60-cm telescope. 
In 2009, we observed only from Jena.
The observations log from 2009 (Jena only) is given in Table 3.
Only in Fig. 6,
we also show a preliminary light curve from the 2010 campaign.
In addition, in Fig. 7, we show a spectrum obtained with the
Jena 90-cm telescope and a fiber-fed spectrograph;
we can take spectra of bright stars (90-cm mirror) also simultaneously
with optical CCD photometry with the small 25-cm telescope,
which can be quite interesting for variable young active stars.
The Tr-37 was observed for three runs in 2010 and will also be
observed for three runs in 2011 and 2012 by most telescopes
of the network. Final results including follow-up
observations will be presented later.

\begin{table}
\begin{tabular}{lllll}
\multicolumn{5}{c}{{\bf Table 3. Observations log form 2009 for Tr-37} (Jena only)} \\ \hline
Date (a)      & begin [UT]     & end [UT]     & filter & exp. [s] (b)  \\ \hline
07/29      & 21:32          & 01:53        & R      & 7790 \\
07/31      & 01:08          & 02:07        & R      & 2660 \\
08/01      & 20:32          & 02:07        & R      & 14650 \\
08/04      & 20:46          & 02:39        & R      & 15260 \\
08/05      & 20:27          & 02:24        & R      & 14160 \\
08/06      & 20:16          & 01:24        & R      & 5840 \\
08/13      & 21:48          & 00:49        & R      & 7070 \\
08/15      & 20:07          & 21:58        & R      & 4980 \\
08/17      & 21:38          & 22:54        & R      & 3430 \\
08/18      & 19:43          & 03:14        & R      & 16720 \\
08/19      & 20:22          & 02:35        & R      & 14220 \\
08/21      & 00:26          & 01:30        & R      & 2380 \\
08/22      & 20:00          & 02:22        & R      & 17260 \\ 
08/23      & 19:32          & 03:05        & R      & 16000 \\
08/24      & 20:50          & 02:23        & R      & 14870 \\
08/26      & 22:07          & 03:01        & R      & 9460 \\
08/27      & 22:18          & 02:56        & R      & 10290 \\
08/29      & 21:37          & 02:26        & R      & 12810 \\
08/30      & 21:07          & 02:00        & R      & 12810 \\
08/31      & 23:16          & 02:54        & R      & 9530 \\
09/05      & 20:33          & 03:05        & R      & 9360 \\
09/07      & 19:00          & 03:27        & R      & 18790 \\
09/08      & 19:08          & 01:52        & R      & 16440 \\
09/09      & 19:00          & 01:50        & R      & 14880 \\
09/18      & 18:58          & 00:08        & R      & 12530 \\
09/19      & 18:20          & 23:29        & R      & 13880 \\
09/21      & 21:40          & 22:05        & BVRI   & 1560 \\
09/22      & 18:49          & 03:54        & R      & 19190 \\
09/25      & 19:54          & 03:16        & R      & 19080 \\
09/26      & 18:04          & 23:07        & R      & 13520 \\
09/27      & 23:07          & 04:17        & R      & 13650 \\
10/20      & 21:27          & 01:50        & R      & 11900 \\
11/06      & 19:59          & 23:38        & R      & 1110 \\ \hline
\end{tabular}
Remarks: (a) Dates (month/day) for the beginning of the observing nights, 
all in the year 2009; (b) total exposure of all images in that night,
split roughly half and half into individual exposures of 10s and 60s.
\end{table}

Basic data reduction of the Jena data from 2009 shown in the figures here
includes bias, dark, flat-field, illumination, and bad pixel corrections.
Relative photometry follows the procedure described in Broeg et al. (2005): 
We compare each star in the FoV with all other stars to investigate
its variability, then we construct
an artifical comparison star, then iterate by giving variable stars less weight;
the final artifical comparison star
is then very constant. We do this not only for one particular program star, as in 
Broeg et al. (2005), but for all stars in the field. For details, see Broeg et al. (2005).
Finally, we also test de-trending (Tamuz, Mazeh, Zucker 2005).

Periodic variability can then be detected by typical period search procedures
like Stringlength, Lomb-Scargle, or Fourier analysis (see e.g. Berndt et al. 2011
for rotation periods of members of Tr-37).
We are currently implementing a Bayesian transit detection routine (Hambaryan et al., in prep.)
to search for the exact type of the signal (planetary transit), even if not strictly periodic,
as in the case of TTV signals or very long periods (so that only one transit is observed).

\begin{figure}
{\includegraphics[angle=270,width=8cm]{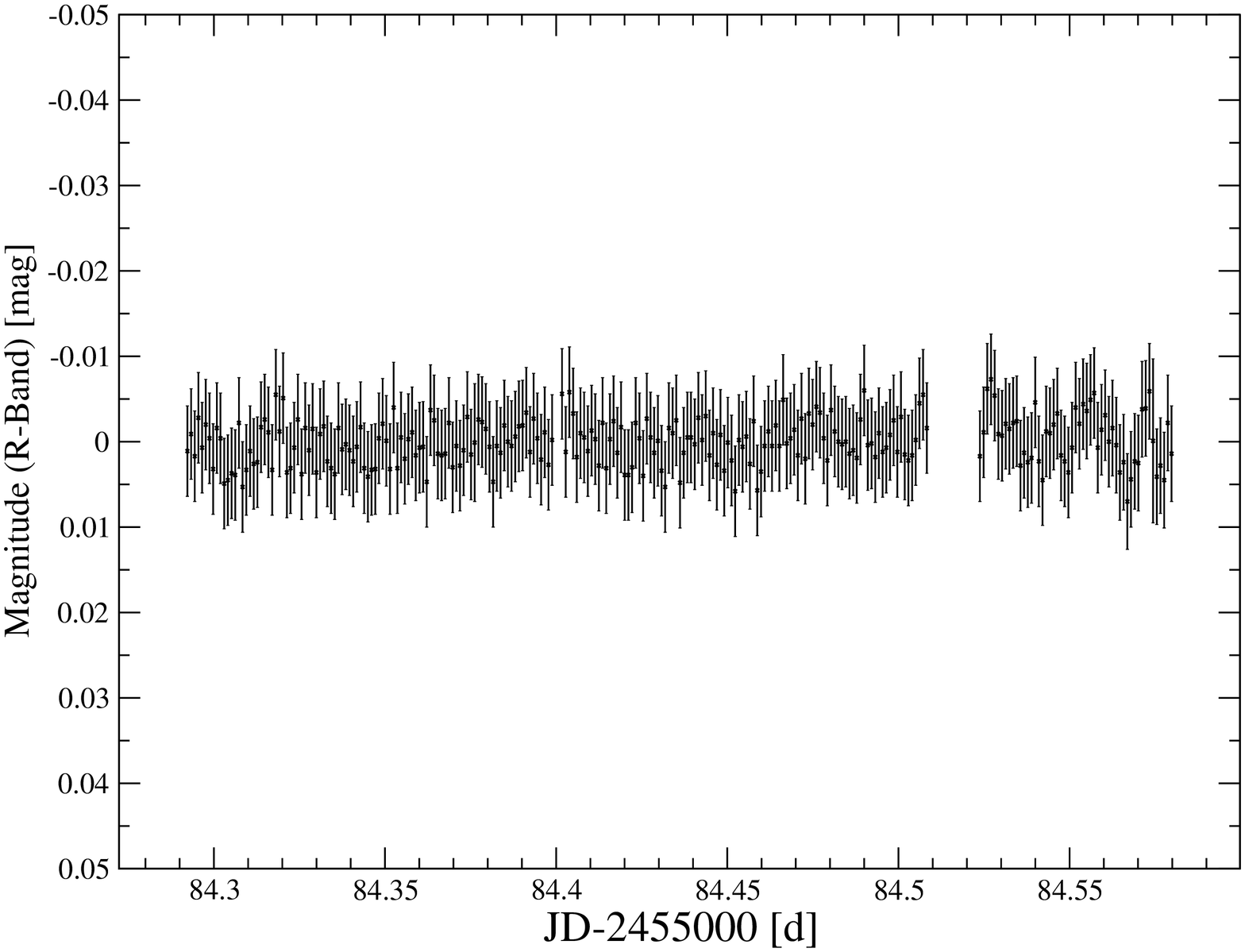}}
\caption{R-band light curve from Jena STK CCD from 2009 Sept 9
for the star MVA 497, spectral type A1 (Sicilia-Aguilar et al. 2005).
It has a proper motion membership probability of 0.92 (Marschall \& van Altena 1987).
We obtained R=12.8 mag and V=12.8 mag, consistent with early A.
This is the most constant member star in the Jena STK data from the best night (lowest
scatter in photometric data of whole ensemble, excluding nights with observations lasting
less than 3 hours). The scatter of this star is $\pm 2.7$ milli-mag. 
The observation in the night ended due to clouds, which arrived at UT 2:00h. 
The first cloud ridge arrived earlier, therefore the scatter at the end of the night is slightly worse.
There were two stars in the FoV with even slightly smaller scatter,
but for those it is not yet known whether they are members or non-members.}
\end{figure}

The first few planetary transit observations with a 25-cm telescope of the
University of Jena Observatory near the village of Gro\ss schwabhausen near Jena
of the objects TrES-1, TrES-2, and XO-1 (Raetz et al. 2008; Va\v{n}ko et al. 2008; Raetz et al. 2009a, 2009b)
show that transits can be detected from the Jena observatory, 
but they do not prove that we can find new transiting planets.
Since early 2009, we also use the 90-cm mirror of the Jena observatory in its
60-cm Schmidt mode ($53^{\prime} \times 53^{\prime}$ FoV) with the
CCD called Schmidt-Teleskop-Kamera (STK), see Mugrauer \& Bertold (2009) for details.
With that telescope and camera, we could also re-detect known planetary
transits with a precision of $\sim 1$ mmag rms scatter and
$\pm 27$s transit timing precision (Maciejewski et al. 2010, 2011a).

We reached 5 mmag rms precision for all un-saturated stars down to R=14.5 mag with
the Jena 90-cm telescope in the 60-cm Schmidt mode (see Fig. 5).
For the other telescopes of the network with similar mirror size
and similar CCD detectors, we can reach a similar precision:
50 mmag rms down to R=17.7 mag for Byurakan 2.6m,
as well as down to R=17.8 mag for Lulin 1m,
down to R=15.4 mag for Swarthmore 0.6m,
and to R=16.9 mag for Xinglong 0.9m; 
also 5 mmag rms down to R=14.9 mag for Byurakan 2.6m,
as well as down to R=15.4 mag for Lulin 1m,
and down to R=13.4 mag for Swarthmore 0.6m:
the data for the other telescopes are still being reduced.

We show a few preliminary results in figures 2 to 6:
Light curves from the 2009 monitoring only at the Jena telescope,
e.g. a 3.5 day rotation period of a classical T Tauri star (Fig. 3) -
but also light curves with obvious gaps (e.g. Fig. 4), 
motivating the collaboration with many other observatories (YETI).
We show
the sensitivity obtained with the Jena 90/60-cm in Fig. 5.
In Fig. 6, we show a preliminary light-curve with data combined from
several telescopes from the 2010 Tr-37 campaign.
We also show
a spectrum obtained with the 90-cm telescope in Jena
(Fig. 7), namely for the brightest member of Tr-37 (HD 206267),
together with a light curve for this star with high time resolution
with 5s exposures (Fig. 8).
In Fig. 9, we show the light curve for an apparently non-variable member star
($\pm 2.7$ mmag).
We also found a few new eclipsing binaries,
both member candidates and field stars. For the member candidates,
low-resolution spectra have been obtained at Calar Alto with CAFOS,
see below, to confirm membership;
high-resolution spectra were obtained at Keck with HIRES to 
measure the masses of the companions (Errmann et al., in prep.).
Since the FIASCO spectrograph at the 90-cm Jena telescope is useful
only down to about 11th mag, we also use the Calar Alto 2.2m/CAFOS 
for low-resolution and the 1.5m Tillinghast Reflector for high-resolution spectra.

The optical spectra of HD 206267A+B (V=5.6 mag), C (V=8.1 mag), and D (V=7.9 mag) were 
obtained with the fiber-fed spectrograph
FIASCO (Mugrauer \& Avila 2009) operated at the Nasmyth port of the 90cm telescope 
of the University Observatory Jena. The data were taken on 2010 June 7,
using three exposures with 600s exposure time each for HD 206267 A+B
and four such 600s exposures for HD 206267 C and D. Standard
calibration was done (dark and flat-field correction, removal of bad
pixels). Flux correction was performed by using a Vega spectrum from the
same night and airmass, standard spectra are from Le Borgne et al. (2003). 
The de-reddened spectrum of HD 206267A+B is consistent with a spectral type of O6 
(as in Simbad and Sota et al. 2011) 
and the spectra of HD 206267 C and D show spectral type B0V (as in Simbad and Sota et al. 2011),
all spectra show H$\alpha$ and He absorption as well as telluric oxygen,
all as expected.

For transit candidates, i.e. stars showing a light curve with
periodic, small, transit-like, flat-bottom dips, we will do the usual
follow-up observations: First, a high-precision photometric light curve
is obtained with a larger telescope to confirm that the dip is consistent
with a planet transit, i.e. shows a flat-bottom light curve. Then, low- to high
resolution spectroscopic reconnaissance spectra are taken, e.g. with the 
2.2. Calar Alto CAFOS spectrograph or the 1.5m Tillinghast 
Reflector at the Whipple Observatory, the latter giving a resolution of 6.5 km/s
down to V=14 mag. If the object is still not found to be
a binary star, then we obtain high-angular resolution follow-up
imaging with Adaptive Optics (AO, large telescope mirror, small PSF, e.g. Subaru) 
to check whether there are background eclipsing binaries in the larger optical PSF
(from imaging photometry with a smaller telescope mirror),
which could mimic a transit-like periodic event.
Then, we can also consider to obtain high-resolution infrared
spectra, to check for background eclipsing binaries even closer
to the target star, i.e. within the AO PSF (not yet done).
If the transit-like event can still only be explained by
a sub-stellar object transiting the star, then the last 
follow-up observation is time-critical high-resolution spectroscopy,
to measure the mass of the transiting companion.
With the RV data, we can then also use the Rossiter McLaughlin effect 
to show that a body of small radius orbits the observed star,
excluding a faint eclipsing binary hidden in the system PSF.

We have done most of those follow-up observations so far
for one transit candidate (Errmann et al., in prep.).
This first candidate was detected in the 2009 data of the
Jena 90/60-cm telescope only. Hence, after combining the data
from all telescopes from the 2010 campaign, we can expect
to detect several more candidates.

We will report results of the photometric monitoring campaigns
and the follow-up observations in the near future.

\section{Summary}

We presented the motivation, observing strategy, target cluster selection, and
first results of the new international multi-site project YETI with its
main goal being the discovery and study of young exoplanets. The photometric precision is 
50 mmag rms down to R=16.5 mag
(5 mmag rms down to R=14.5 mag, both values for Jena 90/60-cm, 
similar brightness limit within about $\pm 1$ mag also for the other telescopes of the network),
i.e. sufficient to detect planetary transits.
We use several telescopes around the world to observe continuously for 24h per
day for several days in order not to miss a transit.

For young transiting exoplanets, we can determine mass and radius, 
hence also the density and possibly the internal structure and composition.
With young exoplanets, one can constrain planet formation models, e.g. whether they
form by gravitational contraction (disk instability) or by core accretion
(nucleated instability), during which 
time-scale and at which separations from the star planets can form.
By comparing different planets found within one cluster,
one can determine the role of stellar parameters like mass on planet formation;
by comparing the planet population between the different clusters, we can study
the impact of environmental conditions on planet formation like metallicity and
density of the cluster.
 
\acknowledgements
All the participating observatories appreciate the logistic and financial
support of their institutions and in particular their technical workshops.
We would like to thank F. Gie\ss ler, H. Gilbert, 
I. H\"ausler, D. Keeley, and S. Fitzpatrick for participating
in some of the observations at the Jena telescope.
RN would like to thank DLN for long-term support and many
good ideas and questions.
RN would like to acknowledge financial support from the Thuringian
government (B 515-07010) for the STK CCD camera used in this project.
RN would like to thank the German National Science Foundation (Deutsche
Forschungsgemeinschaft, DFG) for general support in various projects.
RN, RE, SR, and CA would like to thank DFG for support in the Priority Programme SPP 1385
on the {\em First ten Million years of the Solar System} in projects NE 515 / 34-1,
NE 515 / 33-1, and NE 515 / 35-1.
RN, GM, TP, and MV would like to thank the European Union in the Framework Programme FP6
Marie Curie Transfer of Knowledge project MTKD-CT-2006-042514 for support.
SF thanks the State of Thuringia for a scholarship.
RN, GM, AN, LB, TT, GN, MaMu, SR, and TP would like to thank DAAD PPP-MNiSW 
in project no. 50724260-2010/2011 for support.
RR and UK would like to thank DFG for support in the Priority Programme SPP 1385
in project RE 882/12-1.
AB would like to thank DFG for support in project NE 515 / 32-1.
MoMo would like to thank the Tishreen University in Lattakia, Syria for a scholarship.
TE, NT, MMH, LT, and VVH would like to thank the DFG for support from the SFB-TR 7.
TR would like to thank DFG for support in project NE 515 / 33-1.
CM and KS would like to thank DFG for support in project SCHR 665 / 7-1.
TOBS, ChGi, LT, TE, and TR would like to thank DFG for support in project NE 515 / 30-1.
MS would like to thank DFG for support in project NE 515 / 36-1.
NT would like to thank the Carl-Zeiss-Foundation for a scholarship. 
RN, RE, TP, TOBS, UK, and MS would like to thank DFG for travel support to
Calar Alto runs in projects NE 515 / 38-1 and 39-1.
RN and RE would like to thank DFG for financial support for a
Keck run in project NE 515 / 42-1.
ELNJ and his colleagues from Swarthmore acknowledge support
from the US National Science Foundation grant, AST-0721386.
Osservatorio Cerro Armazones (OCA) is supported as a project 
of the Nordrhein-Westf\"alische Akademie der
Wissenschaften und der K\"unste in the framework of the academy program by
the Federal Republic of Germany and the state Nordrhein-Westfalen.
ZYW is supported by the National Natural Science Foundation of China, No. 10803007.
JB, TP, and MV thank VEGA grants 2/0094/11, 2/0078/10, 2/0074/09.
DPD, DPK, and VSR would like to acknowledge for financial support to the National
Science Foundation of the Bulgarian Ministry of Education and Science (project DO
02-362). We used Simbad, ViziR, 2MASS, and WEBDA.
We would like to thank an anonymous referee for good suggestions.

\newpage%%%%%%%%%%%%%%%%%%%%%%%%%%%%%%%%%%%%%%%%%%%%%%%%%%%%%%

\end{document}